# On the origin of the Raman scattering in heavily boron-doped diamond

**Igor I. Vlasov*,a, Evgeniy A. Ekimovb, Artem A. Basova, Etienne Goovaertsc, Andrey V. Zoteevd**

[a] General Physics Institute, RAS, Vavilov street 38, 119991 Moscow, Russia

[b] Vereshchagin Institute for High Pressure Physics, RAS, 142190 Troitsk, Russia

[c] University of Antwerp, Physics Department, Universiteitsplein 1, 2610 Antwerp, Belgium

[d] Moscow State University, Physics Department, Leninskie gory, 119992 Moscow, Russia

**Abstract**

Isotopic substitution of boron and carbon is applied for the identification of the vibrational modes of heavily boron-doped diamond synthesized by high-pressure high-temperature technique. None of the bands in the Raman spectra are shifting upon $^{10}$B-substitution, whereas shifts to lower frequency are observed for all bands upon $^{13}$C-substitution as compared to a sample with natural isotope abundan[1]cies. These isotopic substitution experiments exclude the hypothesis of boron dimer related normal modes and strongly support the assignment of the previously studied "500 cm$^{-1}$" and "1230 cm$^{-1}$" bands and two weak bands at 1003 cm$^{-1}$ and 1070 cm$^{-1}$ to perturbed diamond lattice phonons, revealing the phonon density of states. A second-order phonon spectrum at combination and overtone frequencies is also identified. A bulk plasmon related mechanism is proposed for the enhancement of the phonon density of states spectrum relative to the zone-center phonons.

## 1. Introduction

Boron is the most efficient dopant in diamond due to the high level of boron solubility in this material. Boron-doped diamond films are increasingly used in the design of electronic devices and in electrochemistry [1]. Lightly boron-doped diamond has p-type conductivity with an activation energy of ca. 0.37 eV [2]. With increasing concentration of boron in diamond beyond 2-5×10$^{20}$ B cm$^{-3}$, the conductivity becomes quasi-metallic [3] at room temperatures. Very recently boron-doped diamond was demonstrated to become superconducting at temperatures in the range of 1-9 K [4-6].

It is evident that high boron concentrations perturb the electronic and vibrational properties of diamond. In particular, Raman spectra of heavily boron-doped diamond (>1×10$^{20}$ B cm$^{-3}$) demonstrate broadening, low-frequency shift, and asymmetry of the line attributed to the zone-center diamond phonons (ZCP), and usually observed at 1332.5 cm$^{-1}$ in undoped diamond. The asymmetry is attributed to the Fano-type interference between the discrete ZCP and the continuum of electronic excitations induced by the presence of the dopant [7]. Besides, a broad spectrum is appearing at lower frequencies than the ZCP line with two most pronounced maxima which we will label "500 cm$^{-1}$" and "1230 cm$^{-1}$" according to their approximate positions previously reported in literature [3,8-10]. Their actual positions depend on the boron concentration in diamond and move to the lower wavenumbers with increasing concentration. The origin of these bands is still under discussion. An early suggestion was that these two maxima may relate to the maxima of the phonon density of states (DOS) of diamond [3,8,9]. The Raman observation of the phonon DOS becomes possible when the wavevector selection rules in the Raman scattering process is relaxed, which occurs when the crystal structure is disturbed by impurities and/or defects [11,12]. For instance, disorder induced features at ca. 500 and 1200 cm$^{-1}$ have been observed in He-ion implanted diamond [13] and in strongly defected CVD diamond films [14], a band around 500 cm$^{-1}$ was found for amorphous carbon deposited by laser ablation in which sp$^3$ bonds were dominating [15]. Likewise, the introduction of a large amount of boron into a diamond lattice would reduce its order, and would allow to observe the diamond phonon DOS. However, it has been found [10,16,17] that the intensities of the "500 cm$^{-1}$" and "1230 cm$^{-1}$" bands decrease compared to that of characteristic diamond line with increasing excitation energy and can not be detected at all when the Raman scattering is observed at 244 nm excitation wavelength [17]. This result challenges the attribution of the extra bands in Raman spectra of heavily boron-doped diamond (HBBD) to vibrational modes of diamond lattice, since their intensities would then be expected to follow the same dependence on the excitation energy for all diamond phonons. This observation triggered the appearance of a new hypothesis for the origin of one of the extra Raman bands. It was suggested that the "500 cm$^{-1}$" band may be associated to the vibrational mode of a boron dimer [18-20] in which boron would preferentially aggregate at high concentrations.

---

[1] Corresponding author. Tel.: +7 495 2321129. E-mail adress: vlasov@nsc.gpi.ru (I.I. Vlasov).

Here we demonstrate the identification of the "500 cm$^{-1}$" and "1230 cm$^{-1}$" vibration modes of heavily boron-doped diamond to only the motion of carbon atoms by isotopic substitution with $^{13}$C and $^{10}$B. Other bands observed in the Raman spectra of HBDD are also identified with diamond phonons. A mechanism based on the bulk plasmon enhancement of Raman scattering in HBBD is suggested for an explanation of the anomalous dependence of the phonon DOS scattering intensity on excitation energy. Note that Raman studies have mostly been performed in HBDD produced under metastable conditions by chemical vapor deposition (CVD) [3,7,9,10,16-18] whereas here we analyze HBDD synthesized in the thermodynamically stable region of diamond using high-pressure high-temperature (HPHT) technique.

## 2. Experimental procedure

The HBDD was synthesized in polycrystalline form in the B-C system at a pressure of 8-9 GPa and a temperature of about 2300-2500 K for 8-10 s. Polycrystalline HBDD with grain size up to 10 μm was produced at the boundary between graphite and a piece of boron. Upon heating under pressure, boron reacts with graphite to form boron carbide and then an eutectic melt, which is a growth medium for graphite-to-diamond transformation [21]. All experiments were carried out under the same conditions at a temperature close to the eutectic melting point. In accordance with current views on the phase diagrams with partial solid solutions, the amount of boron incorporated into the diamond lattice during the synthesis must correspond to its solubility limit. For the synthesis of HBDD from isotopically pure $^{13}$C, amorphous carbon ($^{13}$C 99 at % / $^{12}$C 1 at %, Campro Scientific) was taken instead of graphite in the B-C system. For $^{10}$B -enriched sample preparation an initial amorphous boron powder ($^{10}$B 86 at % / $^{11}$B 14 at %, produced in Russia) was cleaned [22] and sintered into pieces before diamond synthesis. The sample synthesized from carbon and boron (crystalline boron, Aldrich Chem. Co) with natural isotopic compositions is specified as $^{11}$B$^{12}$C. The samples produced with $^{10}$B and $^{13}$C isotopic substitutions are marked as $^{10}$B$^{12}$C and $^{11}$B$^{13}$C, respectively.

A LabRam HR-800 spectrometer with He-Ne laser emitting at 632.8 nm were used for the micro-Raman analysis of these three samples. The 3 mW of laser radiation (measured at the sample) was focused on formless grains of polycrystalline diamond (B-C system) in a spot of about 5 μm in diameter. For each sample Raman spectra were measured in eight different points. The variations in band positions related to inhomogeneous distribution of boron concentration within the sample did not exceed 5 cm$^{-1}$ for all observed bands in the Raman spectra of each sample.

## 3. Results and discussion

Figure 1 shows characteristic Raman spectra of the samples $^{11}$B$^{12}$C, $^{10}$B$^{12}$C, and $^{11}$B$^{13}$C recorded in the range of 200-2800 cm$^{-1}$. The seven most pronounced bands with maxima at 477, 1003, 1070, 1212, 1292, 1705 and 2415 cm$^{-1}$ are seen in the spectrum of the sample $^{11}$B$^{12}$C. The presence of $^{10}$B instead of natural boron in the sample $^{10}$B$^{12}$C doesn't noticeably change the positions of the main bands compare to those observed in the spectrum of the $^{11}$B$^{12}$C sample.

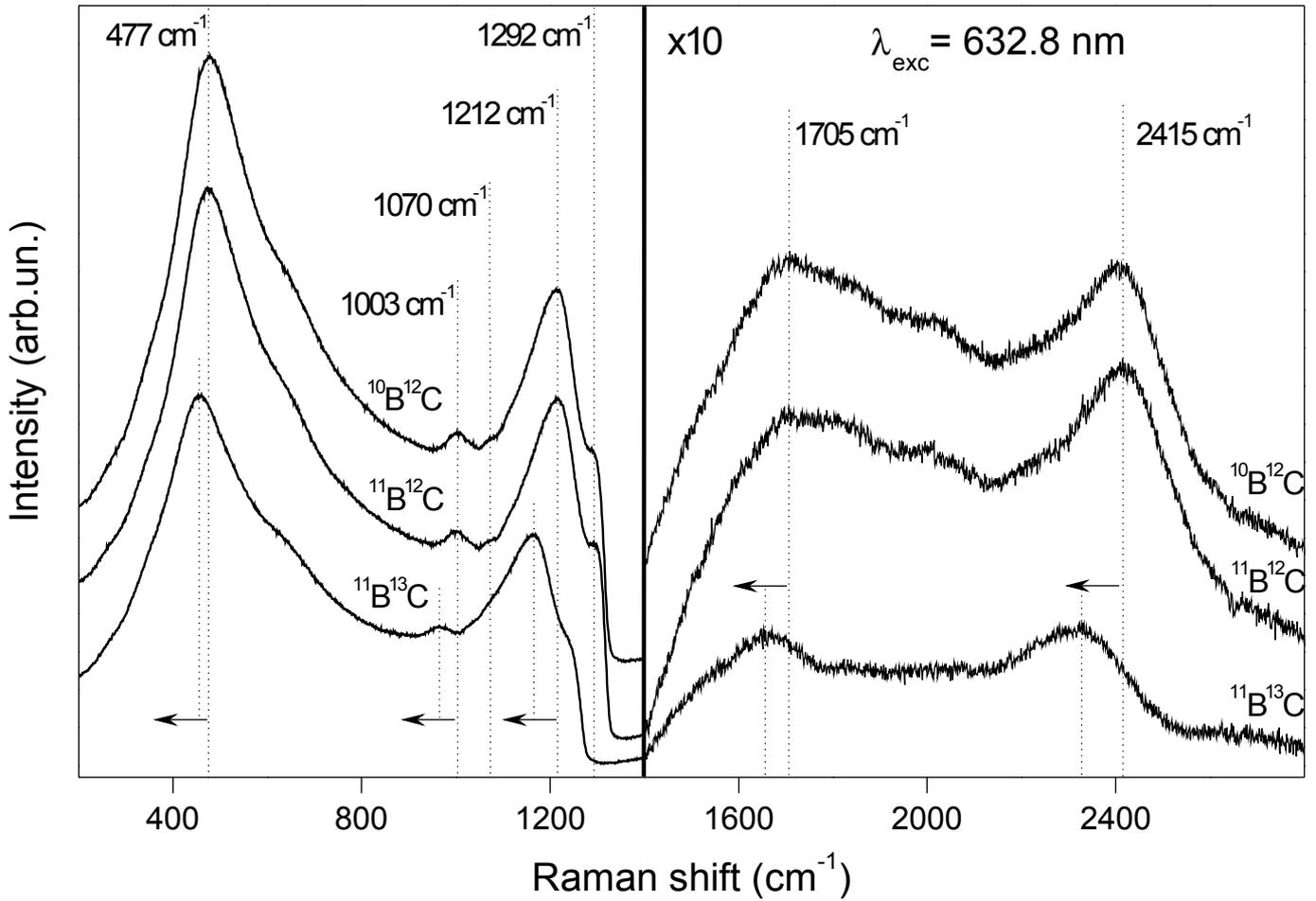

**Figure 1.** The Raman spectra of the $^{10}B^{12}C$, $^{11}B^{12}C$ and $^{11}B^{13}C$ samples recorded at 632.8 nm excitation wavelength. The spectral region of 1400-2800 cm$^{-1}$ in which the second order Raman scattering of diamond is observed, was magnified by factor 10. The arrows show the direction of the line shifts at $^{13}C$ isotope substitution.

When $^{13}C$ is substituted for $^{12}C$ the positions of all the bands are shifted by a factor of 0.95-0.97 which is in good agreement with the expected shift for only carbon related vibration modes, which is estimated as the square root of the ratio of the $^{12}C$ to $^{13}C$ masses (≈0.96). Therefore, we ascribe the 477 cm$^{-1}$ ($v^1_{HBDD}$), 1003 cm$^{-1}$ ($v^2_{HBDD}$), 1070 cm$^{-1}$ ($v^3_{HBDD}$) and 1212 cm$^{-1}$ ($v^4_{HBDD}$) bands to vibration modes in maxima of the phonon DOS of diamond. According to a calculation of lattice dynamical properties of diamond [23] and experimental data [24] these four modes would correspond to TA, LA, TO, LO phonons of diamond, respectively (T – transversal, L – longitudinal, A – acoustic, and O – optic). Note that 477cm$^{-1}$ and 1212 cm$^{-1}$ are the actual positions of the "500 cm$^{-1}$" and "1230 cm$^{-1}$" bands for our diamond samples with extremely high concentration of boron. The 1292 cm$^{-1}$ band belongs to zone-center phonons of HBDD. The two high frequency bands in the Raman spectra, at 1705 cm$^{-1}$ and 2415 cm$^{-1}$ in both the $^{11}B^{12}C$ and $^{10}B^{12}C$ samples, can be ascribed to the combination $v^1_{HBDD} + v^4_{HBDD}$ and to the first overtone of $v^4_{HBDD}$, respectively. The combinations of the $\nu^1_{HBDD}$ and $\nu^4_{HBDD}$ with $\nu^2_{HBDD}$ and $\nu^3_{HBDD}$ seem be too weak in intensity to be distinguished in the Raman spectra of Fig.1. The overtone of $\nu^1_{HBDD}$ is supposed to remain unresolved, because its expected location falls in the minimum between the "500 cm$^{-1}$" and "1230 cm$^{-1}$" bands, just below the 1003 cm$^{-1}$ peak [25]. It is worth pointing out that no Raman band is observed around 1580 cm$^{-1}$, which demonstrates that graphitic-like regions do not occure in these HBBD samples.

We suggest the following mechanism of Raman scattering in HBDD as an explanation of the anomalous dependence [10,16,17] of the phonon DOS scattering intensity on excitation energy. The observation of this spectrum simultaneously with the ZCP line results from an inhomogeneous distribution of the boron substitutional impurity within a crystal. In relatively small regions of the crystal, a very high boron concentration introduces strong substitutional disorder (SD) which is breaking the translation symmetry of the crystal lattice and, as a consequence, the wave-vector conservation rule for Raman scattering. As a result, phonons with a large range of wave vectors in the Brillouin zone are excited and the resultant Raman spectra tends to be broad and reflect the phonon density of states. In these SD regions the carrier density is expected to be $> 10^{21}$ cm$^{-3}$. The existence of bulk plasmons in heavily doped semiconductors was discussed as early as 1965 [26]. Their resonant frequency is proportional to the square root of the carrier density. While plasmon absorption in metals (the carrier density is $\sim 10^{23}$ cm$^{-3}$) is usually observed in the UV region, this occurs at lower frequencies even in heavily doped semiconductors, because of the lower carrier density [26], eventually shifting to the near infrared (NIR) for very high concentrations. In these circumstances, we expect that the bulk plasmons in the SD regions give rise to strong enhancement of the local electromagnetic field leading to intense Raman scattering from the diamond phonon DOS. This effect is similar to surface enhanced Raman scattering (SERS) [27] in semiconductors covered with metallic nanoparticles, and could in analogy be called bulk enhanced Raman scattering (BERS). Here the role of metallic nanoparticles is played by the SD regions with highest concentration of the free carriers. When the Raman excitation is varied from the visible to the NIR, approaching the plasmon resonance, this will enhance the Raman scattering from the phonon DOS in the SD regions.

In the remaining crystal volume the substitutional disorder is weaker, and the first-order Raman activity is largely restricted to the phonons with almost zero wave vector. The decrease of the enhancement factor for BERS when the excitation photon energy moves away from the plasmon range is a first reason for the disappearance of the broad phonon spectrum for UV excitation, leaving only the narrow ZCP band [17]. However, a second reason for this sudden disappearance could be the strong inherent optical absorption of diamond in the UV. As a result, the observed Raman scattering is originating from a sub-surface layer of only a few tens of nanometers. This layer in semiconductors is in general depleted from free carriers [28], due to compensation of the dopants by surface states in the band gap, and therefore no plasmon enhancement of Raman scattering from this layer can be expected. Note that the absence of the broad band Raman scattering under UV excitation reveals that the SD regions, with strongly broken translation symmetry and very high carrier densities, represent only a very small volume fraction in the HBDD diamond crystals.

4. Summary

The Raman study of heavily boron-doped diamond synthesized by high-pressure high-temperature technique using $^{10}$B and $^{13}$C enriched materials has provided new insight into the origin of the features observed in Raman spectra of the HBDD. Namely, the "500 cm$^{-1}$" and "1230 cm$^{-1}$" bands widely discussed in literature were unequivocally identified with vibrational modes of diamond. Besides, four weak bands observed at lower (1003 cm$^{-1}$ and 1070 cm$^{-1}$) and higher (1705 cm$^{-1}$ and 2415 cm$^{-1}$) frequencies as compared to the ZCP position, were assigned, respectively, to first and second order Raman scattering from the diamond phonons.

Acknowledgements

This work was partially supported by grant No. 05-02-17287 of the Russian Foundation for Basic Research, INTAS Young Scientist Fellowship Ref. Nr. 06-1000014-5934 (A.A.B.), and by grant No. BIL05/RU/37.